\documentclass[twocolumn,amsmath,amssymb,showpacs,aps]{revtex4}
\usepackage{graphicx}
\usepackage{dcolumn}
\usepackage{bm}

\begin{document}
\title{Operationally Invariant Measure of the Distance between Quantum States by Complementary Measurements}

\author{Jinhyoung Lee}

\email{jlee@am.qub.ac.uk}

\author{M. S. Kim}

\affiliation{School of Mathematics and Physics, The Queen's
  University, Belfast, BT7 1NN, United Kingdom}

\author{\v{C}aslav Brukner}

\affiliation{Institut f\"{u}r Experimentalphysik, Universit\"{a}t Wien,
  Boltzmanngasse 5, A-1090 Wien, Austria}

\date{\today}

\begin{abstract}
  We propose an operational measure of distance of two quantum states,
  which conversely tells us their closeness.  This is defined as a sum
  of differences in partial knowledge over a complete set of mutually
  complementary measurements for the two states. It is shown that the
  measure is operationally invariant and it is equivalent to the
  Hilbert-Schmidt distance. The operational measure of distance provides
  a remarkable interpretation of the information distance between
  quantum states.
\end{abstract}


\maketitle

{\em Introduction} -- Mathematical formulations of all the fundamental
physical theories are based on the concept of an abstract space.  The
structure of the space and the theories is defined by its metric. For
example, the Minkowski metric defines the mathematical structure of the
special theory of relativity and the Rieman metric defines the structure
of the general theory of relativity. In quantum mechanics the
Hilbert-Schmidt distance may be the natural metric of the Hilbert space.
What are the fundamental laws which dictate the metrics in physical
theories? This question is investigated in this paper for the case of
quantum theory and the Hilbert-Schmidt distance.

When two quantum states are given, what do we do to measure how close
they are? This is an important issue in various investigations of
quantum mechanics. For example, we need to measure how close the
teleported state is to the original state in order to check the
credibility of the quantum teleportation protocol. Other examples appear
in quantum cloning, quantum state reconstruction, and practical quantum
gate operation \cite{Nielsen00}. We need a measure of closeness,
depending on the kind of information process involved.  In particular,
two measures have been applied to the wide realm of quantum information
processing: Fidelity \cite{Jozsa94} and Hilbert-Schmidt distance
\cite{Buzek96}. These measures are equivalent to each other if the
systems are in pure states.

The fidelity, $F=|\langle \psi| \phi\rangle|^2$, is the transition
probability between two pure states, $|\psi\rangle$ and $|\phi\rangle$.
When the fidelity is extended to incorporate mixed states
\cite{Jozsa94}, its interpretation becomes vague in an operational
perspective. Instead, the fidelity may be indirectly interpreted in
terms of statistical distance or ``statistical distinguishability'' in
the measurement that optimally resolves neighboring density operators
\cite{Braunstein94}. On the other hand, the Hilbert-Schmidt distance is
a metric defined on the space of operators. It is unclear how to impose
an operational interpretation on the Hilbert-Schmidt distance.  Another
possible measure of closeness is quantum relative entropy which has also
been proposed as a candidate for a measure of entanglement
\cite{Vedral97,Umegaki62}.

A quantum state is a representation of our knowledge on individual
outcomes in future experiments \cite{Schrodinger35}. We can, then
intuitively, say that the difference between this knowledge for two
quantum states measures how much the two states are ``close to each
other'' with respect to the future predictions. Bohr \cite{Bohr58}
remarked that ``...phenomena under different experimental conditions,
must be termed complementary in the sense that each is well defined and
that together they exhaust all definable knowledge about the object
concerned.'' This suggests that the closeness of two quantum states
should be defined with regard to a complete set of mutually
complementary measurements. We require that such a measure of closeness
between two states is invariant under the specific choice of a complete
set of mutually complementary measurements.

In this paper we introduce a measure of distance between two quantum
states, which conversely tells us the closeness. The measure of distance
is operationally defined as a sum of the differences in partial
knowledge over a complete set of mutually complementary ``unbiased''
measurements.  The measure has several remarkable properties.  a) The
measure is operationally invariant: It is uniquely defined, being
independent of the specific choice of a complete set of complementary
measurements.  b) The measure is equivalent to the Hilbert-Schmidt
distance. c) The operational measure of distance can be interpreted as
an {\em information distance} between two quantum states.  In addition,
the fact that the operational measure is equivalent to the
Hilbert-Schmidt distance, suggests that the intrinsic structure of
Hilbert space reflects information-theoretical foundations of quantum
theory.

{\em Mutually complementary measurement} -- Two measurements are
mutually complementary if precise knowledge in one of them implies that
all possible outcomes in the other are equally probable
\cite{Englert92}.  Consider a nondegenerate and orthogonal measurement
$A$ represented by a set of eigen projectors $\{\hat{A}_i\}$.  Suppose a
quantum system in $d$-dimensional Hilbert space is prepared in such a
state that the outcome in the measurement $A$ can be predicted with
certainty, for instance, the system's density operator is given by
$\hat{\rho}=\hat{A}_i$. Let $B$ be another nondegenerate and orthogonal
measurement represented by a set of eigen projectors $\{\hat{B}_i\}$.
For a given state (density operator) $\hat{A}_i$, the probability of an
outcome $j$ in the measurement $B$ is given by $p_{j|i} = \mbox{Tr}
\hat{B}_j \hat{A}_i$. The measurement $B$ is mutually complementary to
$A$ if outcomes of measurement $B$ are equally probable:
\begin{eqnarray}
  \label{eq:cmcm}
  p_{j|i} = \frac{1}{d}, ~~~~~~\forall i, j=1,2,..,d.
\end{eqnarray}
A set of complementary measurements is a complete set if the measurement
operators can expand any density operators on the Hilbert space (see
Eq.~(\ref{eq:ocrcbv})). For a spin-1/2 system such a complete set of
complementary measurements is associated with three Pauli spin operators
$\{\hat{\sigma}_x, \hat{\sigma}_y, \hat{\sigma}_z\}$.

{\em Definition of operational distance} -- Consider two quantum systems
of the $d$-dimensional Hilbert space. In order to indicate how close
their density operators $\rho_1$ and $\rho_2$ are to each other, we
consider a complete set of mutually complementary measurements
$M=\{m_\alpha\}$ which are {\em nondegenerate and orthogonal}. Consider
a measuring device set up with the observable for measurement $m_\alpha$
and let $\{\hat{m}_{\alpha,i}\}$ be the set of the eigen operators and
$\{p_{\alpha,i}=\mbox{Tr}(\hat{m}_{\alpha,i} \hat{\rho})\}$ be the set
of probabilities corresponding to the outcomes for a given density
operator $\hat{\rho}$.
The measurement is performed independently and equivalently for each
quantum system and its probability vector is denoted as
$\vec{p}_\alpha(S)$ for system $S$. The distance of the two probability
vectors, $\vec{p}_\alpha(1)$ and $\vec{p}_\alpha(2)$, is defined as,
\begin{eqnarray}
\label{eq:indclo}
D_\alpha(\hat{\rho}_1,\hat{\rho}_2)=|\vec{p}_\alpha(1)-\vec{p}_\alpha(2)|^2.
\end{eqnarray}
The distance $D_\alpha$ is called a single operational distance for
measurement $m_\alpha$ among a complete set of mutually complementary
measurements. The total operational distance may be defined by summing
single operational distances over the complete set of complementary
measurements:
\begin{eqnarray}
\label{eq:totclo}
  D_{total}(\hat{\rho}_1,\hat{\rho}_2) = \sum_{\alpha} D_\alpha(\hat{\rho}_1,\hat{\rho}_2).
\end{eqnarray}

{\em Number of complementary measurements} -- Consider a Hilbert-Schmidt
space ${\cal B}$ of bound operators for a system $S$ in $d$-dimensional
Hilbert space ${\cal H}_d$, in which the inner product of $\hat{A},
\hat{B} \in {\cal B}$ is defined as \cite{Reed80}
\begin{eqnarray}
  \label{eq:iphs}
  (\hat{A}|\hat{B}) = \mbox{Tr} \hat{A}^\dagger \hat{B}.
\end{eqnarray}
The space ${\cal B}$ forms a $d^2$-dimensional vector space where
each element is an operator.  A Hilbert-Schmidt norm of $\hat{A}$
is given by $||\hat{A}||^2 \equiv (\hat{A}|\hat{A})$.  For
operator space ${\cal
  B}$, one may choose a complete orthogonal basis set in terms of
Hermitian operators, $B_o=\{\hat{\lambda}_\alpha,~\mbox{for}~
\alpha=0,1,...,d^2-1\}$, such that $\hat{\lambda}_0 = \openone$ and
$(\hat{\lambda}_\alpha|\hat{\lambda}_\beta) = d \delta_{\alpha\beta}$.
The orthogonality implies that each $\hat{\lambda}_\alpha$ for $\alpha
\ne 0$ is traceless: $\mbox{Tr}\hat{\lambda}_\alpha = 0$.

A Hermitian operator $\hat{H}$ and a density operator $\hat{\rho}$ of
$S$ are represented by the observable basis set $B_o$ as
\begin{eqnarray}
  \label{eq:rhobob}
  \hat{H} &=& \frac{h_0}{d} \openone + \frac{1}{d}\sum_{\alpha=1}^{d^2-1} h_\alpha
  \hat{\lambda}_\alpha, \\
  \label{eq:rdobob}
  \hat{\rho} &=& \frac{1}{d} \openone + \frac{1}{d}\sum_{\alpha=1}^{d^2-1} \rho_\alpha
  \hat{\lambda}_\alpha,
\end{eqnarray}
where $h_0 = \mbox{Tr} \hat{H}$, $h_\alpha = \mbox{Tr}
\hat{\lambda}_\alpha \hat{H}$, and $\rho_\alpha = \mbox{Tr}
\hat{\lambda}_\alpha \hat{\rho}$. Here $\rho_0=1$ due to the unit trace
of a density operator. In particular we call $\vec{\rho} \equiv (\rho_1,
\rho_2,..., \rho_{d^2-1})$ {\em a generalized Bloch vector}. Because
$\mbox{Tr} \hat{\rho}^2 \le 1$, the norm of $\vec{\rho}$ is upper
bounded: $|\vec{\rho}|^2 \le d-1$.  If $\hat{\rho}$ is pure,
$|\vec{\rho}|^2 = d-1$. The generalized Bloch vectors stay within a
Bloch sphere $S_B$ of radius $\sqrt{d-1}$.  However, not all generalized
Bloch vectors within $S_B$ correspond to density operators, implying
there is no one-to-one correspondence between density operators and
generalized Bloch vectors within the Bloch sphere $S_B$. In fact, the
set of Bloch vectors specifying density operators is restricted by the
positivity of density operators such that a given density operator
$\hat{\rho}$ should hold
\begin{eqnarray}
  \label{eq:podo}
  (\hat{\rho}|\hat{\sigma}) \ge 0 \Leftrightarrow
  \vec{\rho} \cdot \vec{\sigma} \ge -1,
\end{eqnarray}
for any pure density operator $\hat{\sigma}$ with
$|\vec{\sigma}|^2=d-1$.

We shall derive a condition of mutual complementarity with respect to
generalized Bloch vectors. Consider two measurements $A$ and $B$ of
$\{\hat{A}_i\}$ and $\{\hat{B_i}\}$, respectively.  The orthogonality
and the completeness relation of $\{\hat{A}_i\}$ raise relations among
their generalized Bloch vectors $\{\vec{a}_i\}$ as, noting that
$\hat{A}_i$ has unit trace,
\begin{eqnarray}
  \label{eq:ocbv}
  \mbox{Tr}\hat{A}_i \hat{A}_j = \delta_{ij} & \rightarrow &
  \vec{a}_i\cdot\vec{a}_j = d \delta_{ij} -1, \\
  \label{eq:crbv}
  \sum_{i=1}^d\hat{A}_i = \openone & \rightarrow & \sum_{i=1}^{d} \vec{a}_i=\vec{0},
\end{eqnarray}
where $\vec{0}$ is a null vector.  Similar relations hold for the
generalized Bloch vectors $\{\vec{b}_i\}$ of the measurement
$B$.  The condition (\ref{eq:cmcm}) of mutual complementarity
between $A$ and $B$ is now written as
\begin{eqnarray}
  \label{eq:cmcmgbv}
    \vec{a}_i\cdot\vec{b}_j=0, ~~~~~~\forall i, j=1,2,..,d.
\end{eqnarray}
This condition implies that the subspace spanned by $\{\vec{a}_i\}$ is
orthogonal to that by $\{\vec{b}_i\}$ within $S_B$ when $A$ and $B$ are
mutually complementary.

The subspace spanned by $\{\vec{a}_i\}$ is $(d-1)$-dimensional due to
the constraints in Eqs.~(\ref{eq:ocbv}) and (\ref{eq:crbv}). Further,
the set satisfies an over-completeness relation in the subspace as
\begin{eqnarray}
  \label{eq:ocrbv}
  \frac{1}{d} \sum_{i=1}^d \vec{a}_i\vec{a}_i = \openone_{d-1}
\end{eqnarray}
where $\vec{a}_i\vec{a}_j$ is a tensor product of two vectors
$\vec{a}_i$ and $\vec{a}_j$ and $\openone_{d-1}$ is an identity matrix
in the subspace.  Noting that the Bloch space is $(d^2-1)$-dimensional,
it can be divided into $(d+1)$ subspaces in $(d-1)$ dimension. {\em For
  the $d$-dimensional Hilbert space ${\cal H}_d$, there are thus $(d+1)$
  measurements that are mutually complementary and they form a complete
  set of complementary measurements}. We note here that, even though a
pair of mutually complementary measurements always exists, the existence
of a complete set need to be investigated in the virtue of the
condition~(\ref{eq:podo}) and was constructed explicitly for $d$ being a
prime or a power of a prime number \cite{Wootters89}. This finding does
not, however, exclude a possibility to find a complete set of mutually
complementary measurements for other dimensions. To avoid any confusion,
we are concerned with quantum systems in dimensions of prime numbers and
their powers.

We present a nontrivial example of $d$ being a prime number for a
complete set of mutually complementary measurements \cite{Wootters89}.
Consider a measurement which is represented by a basis set
$\{|\phi^0_j\rangle=|j\rangle\}$ and further $d$ measurements, among
which the $\alpha$-th measurement is represented by the basis vectors
\begin{eqnarray}
  \label{eq:cmcmsb}
  |\phi^\alpha_j\rangle = \frac{1}{\sqrt{d}} \sum_{k=1}^d\exp
   \left[\left(2\pi i /d \right) \left(\alpha k^2+j k\right) \right]
   |k\rangle, 
\end{eqnarray}
for $j=1,2,...,d$. One can verify that each of these $(d+1)$ basis
sets is orthonormal and that all the basis sets are mutually
complementary.

{\em Equivalence to Hilbert-Schmidt distance} -- We shall derive one of
the main results that the total operational distance is equivalent to
the Hilbert-Schmidt distance. Let $M$ be a complete set of $(d+1)$
complementary measurements. For $m_\alpha \in M$ with eigen projectors
$\{\hat{m}_{\alpha,i}\}$, let $\vec{m}_{\alpha,i}$ be the generalized
Bloch vector of $\hat{m}_{\alpha,i}$. Because the set
$\{\vec{m}_{\alpha,i}\}$ is over-completed in the Bloch space due to
Eq.~(\ref{eq:ocrbv}),
\begin{eqnarray}
  \label{eq:ocrcbv}
  \frac{1}{d} \sum_{\alpha=1}^{d+1} \sum_{i=1}^d \vec{m}_{\alpha,i}
  \vec{m}_{\alpha,i} = \sum_{\alpha=1}^{d+1} \openone^\alpha_{d-1} = \openone_{d^2-1},
\end{eqnarray}
where $\openone^\alpha_{d-1}$ is a projection matrix onto the
$\alpha$-th subspace and $\openone_{d^2-1}$ is an identity matrix in the
Bloch space.

We obtain a single operational distance explicitly by complementary
measurement $m_\alpha \in M$ and the total operational distance for
given two density operators $\hat{\rho}_1$ and $\hat{\rho}_2$. For a
measurement $m_\alpha \in M$, the single operational distance is given
by Eq.~(\ref{eq:indclo}) as
\begin{eqnarray}
  D_\alpha(\hat{\rho}_1,\hat{\rho}_2) &=& \frac{1}{d^2} \sum_{i=1}^{d} \left|
    \vec{m}_{\alpha,i} \cdot [\vec{\rho}(1) - \vec{\rho}(2)]\right|^2
\end{eqnarray}
where $\vec{\rho}(S)$ is a generalized Bloch vector for the system $S$.
Summing up the single operational distances over the complete set of
complementary measurements, the total distance is obtained by
Eq.~(\ref{eq:totclo}) as
\begin{eqnarray}
  \label{eq:tc}
  D_{total}(\hat{\rho}_1,\hat{\rho}_2) = ||\hat{\rho}_1 - \hat{\rho}_2||^2,
\end{eqnarray}
where we have used the completeness relation~(\ref{eq:ocrcbv}).

We remark some properties of the total distance $D_{total}$.  First, the
total distance is invariant to the specific choice of a complete set of
complementary measurements. In fact, in deriving Eq.~(\ref{eq:tc}), no
particular set of complementary measurements has been chosen. Second,
the total distance is equal to the Hilbert-Schmidt distance of the two
operators $\hat{\rho}_1$ and $\hat{\rho}_2$ in the Hilbert-Schmidt space
${\cal B}$. Third, the total distance is bounded:
\begin{eqnarray}
  \label{eq:bctc}
  0 \le D_{total} \le 2.
\end{eqnarray}
where the bound values 0 and 2 are obvious as shown later in
Eq.~(\ref{eq:tcfps}).

{\em Relation between operational distance and information content} --
Brukner and Zeilinger \cite{Brukner99} introduced the total information
content of a quantum system in the density operator $\hat{\rho}$ and it
was successfully applied for entanglement teleportation \cite{Lee00},
state estimation \cite{Rehacek02}, and a criterion for the violation of
Bell's inequalities \cite{Brukner01}.  Their measure can be written as
\begin{eqnarray}
  \label{eq:mqi}
  I(\hat{\rho}) = N ||\hat{\rho} - \hat{\rho}_r||^2
\end{eqnarray}
where $N$ is a normalization factor and $\hat{\rho}_r=\frac{1}{d}
\openone$ is a completely random state. Comparing Eq.~(\ref{eq:mqi})
with Eq.~(\ref{eq:tc}), the total information content $I(\hat{\rho})$
can be described in terms of the total operational distance
$D_{total}(\hat{\rho},\hat{\rho}_r)$ such that $I(\hat{\rho})$ indicates
the distance of the quantum state $\hat{\rho}$ from the completely
random state $\hat{\rho}_r$. The more information a density operator
$\hat{\rho}$ contains, the further it is away from $\hat{\rho}_r$.
Reciprocally, the total operational distance between two density
operators $\hat{\rho}_1$ and $\hat{\rho}_2$,
$D_{total}(\hat{\rho}_1,\hat{\rho}_2)$, describes a difference in their
information contents. These results imply that the total operational
distance can be interpreted as an {\em information distance} between two
quantum states.

{\em Comparison with fidelity} -- In the following discussion we compare
the total operational distance with the fidelity. The fidelity has been
commonly employed for a measure of closeness in quantum information
processing. The fidelity $F$ is defined by \cite{Jozsa94}
\begin{eqnarray}
  \label{eq:sf}
  F(\hat{\rho}_1,\hat{\rho}_2)=\left(\mbox{Tr} \sqrt{\sqrt{\hat{\rho}_1}
  \hat{\rho}_2 \sqrt{\hat{\rho}_1}}\right)^2,
\end{eqnarray}
for two density operators $\hat{\rho_1}$ and $\hat{\rho}_2$. The
fidelity is bounded by its definition: $0 \le F \le 1$. The two density
operators are exactly same if $F=1$ and they are completely different if
$F=0$. Note for the total operational distance that two operators are
equal if $D=0$ and they are completely different if $D=2$.

One may compare a set of test density operators $\{\hat{\rho}\}$ to a
reference density operator $\hat{\sigma}$ so as to find out which
density operator is the closest to $\hat{\sigma}$. For that purpose, let
us denote the fidelity as $F_{\hat{\sigma}}(\hat{\rho}) \equiv
F(\hat{\sigma},\hat{\rho})$ for a reference density operator
$\hat{\sigma}$.  Similarly, $D_{\hat{\sigma}}(\hat{\rho}) \equiv
D_{total}(\hat{\sigma},\hat{\rho})$.

Consider a measure $M(\vec{q})$ of physical quantities $\{\vec{q}\}$.
The measure $M$ establishes the ordering of physical quantities such
that $M(\vec{q}_1) \le M(\vec{q}_2) \le ... $.  Another measure
$N(\vec{q})$ of physical quantities is {\em equivalent} to $M(\vec{q})$
if $N$ is a monotonic function of $M$, in other words, if the ordering
is either preserved ($N(\vec{q}_1) \le N(\vec{q}_2) \le ...  $) or
completely reversed ($N(\vec{q}_1) \ge N(\vec{q}_2) \ge ...  $).  The
fidelity $F_{\hat{\sigma}}$ is equivalent to the total operational
distance $D_{\hat{\sigma}}$ over a set of test density operators $T$ for
a reference $\hat{\sigma}$ if, for each pair of two test density
operators $\hat{\rho}_1, \hat{\rho}_2 \in T$,
\begin{eqnarray}
  \label{eq:tmord}
  F_{\hat{\sigma}}(\hat{\rho}_1) \le F_{\hat{\sigma}}(\hat{\rho}_2)
  \Leftrightarrow D_{\hat{\sigma}}(\hat{\rho}_1) \ge D_{\hat{\sigma}}(\hat{\rho}_2).
\end{eqnarray}
Further, the fidelity $F$ is equivalent to the total distance $D$ over a
set of test density operators $T$ for a set of reference density
operators $S$ if $F_{\hat{\sigma}}$ is equivalent to $D_{\hat{\sigma}}$
for any reference $\hat{\sigma} \in S$.

If the set of test and reference density operators are confined to pure
states, the fidelity is equivalent to the total operational distance.
Note that for a set of pure states the fidelity is given by the
Hilbert-Schmidt inner product of two density operators
$\hat{\sigma}=|\sigma\rangle\langle\sigma|$ and
$\hat{\rho}=|\rho\rangle\langle\rho|$ in Eq.~(\ref{eq:iphs}):
\begin{eqnarray}
  \label{eq:sthsip}
  F_{\hat{\sigma}}(\hat{\rho}) = \mbox{Tr} \hat{\sigma} \hat{\rho} =
  (\hat{\sigma}|\hat{\rho}).
\end{eqnarray}
Further,
\begin{eqnarray}
  \label{eq:tcfps}
  D_{\hat{\sigma}}(\hat{\rho}) 
  =  P({\hat{\sigma}}) + P({\hat{\rho}}) - 2 F_{\hat{\sigma}}(\hat{\rho}),
\end{eqnarray}
where $P({\hat{\rho}})=||\hat{\rho}||^2$ is the purity of
$\hat{\rho}$.  As $P({\hat{\sigma}})=P({\hat{\rho}})=1$, it is
clear in Eq.~(\ref{eq:tcfps}) that the total operational distance
is monotonic function of the fidelity.

{\em The total operational distance is, however, inequivalent to the
  fidelity as general mixed states are concerned.} For simplicity, let a
reference be a pure state, $\hat{\sigma}=|\sigma\rangle\langle\sigma|$,
and let $P(\hat{\rho})$ be the purity of a test state $\hat{\rho}$. In
the case the fidelity $F_{\sigma}(\rho)$ is written as in
Eq.~(\ref{eq:sthsip}) and the total distance is given as in
Eq.~(\ref{eq:tcfps}) with $P(\hat{\sigma})=1$. Now the total distance is
not just a function of $F_{\hat{\sigma}}(\hat{\rho})$ but also a
function of $P(\hat{\rho})$. As $F$ and $P$ are independent quantities
over a set of test states $\{\hat{\rho}\}$, the
equivalence~(\ref{eq:tmord}) no longer holds.

{\em Quantum tomography and operational distance for a qubit} -- As an
example to obtain the operational distance in an experiment, we consider
quantum tomography on light fields which are an ensemble of polarization
degrees of freedom \cite{Leonhardt97}. The tomographic experiment
obtains Stokes parameters by four intensity measurements \cite{James01}
a) with a filter that transmits 50\% of the incident radiation
regardless of its polarization, b) with a polarizer that transmits only
horizontally polarized light, c) with a polarizer that transmits only
light polarized at $45^\circ$ to the horizontal axis, and d) with a
polarizer that transmits only right-circularly polarized light.  The
latter three cases are propositions for a complete set of mutually
complementary measurements for a polarization qubit. Thus the
operational distance of two ensembles of light fields may be estimated
using the tomography setup.

% In various experimental circumstances subtle treatments such as
% ``maximum likelihood'' tomographic approach are required to estimate a
% legitimate density operator as measuring devices may introduce noise
% \cite{James01}. On the other hand, the operational distance may be
% estimated by a linear tomography method. Assuming that the noise
% introduced to the probability vectors depends only on the measuring
% devices, the contribution by the noise is eliminated (or considerably
% reduced) in Eq.~(\ref{eq:indclo}) due to the subtracting nature of the
% operational distance.

In summary, we proposed a measure to find how close two quantum states
are.  This is operationally defined with respect to a complete set of
mutually complementary measurements.  It was shown that the operational
measure is equivalent to the Hilbert-Schmidt distance, which implies
that our result can also be understood as an operational determination
of the Hilbert-Schmidt distance. The measure provides a remarkable
interpretation as an information distance between quantum states. The
comparison with the fidelity shows that the measure is not necessarily
equivalent to the fidelity.

We thank A. Zeilinger and I. B. Kim for stimulating discussions. This
work is supported by the UK Engineering and Physical Sciences Research
Council for financial support through GR/S14023/01.  \v{C}.B. was
supported by the Austrian FWF Project No. F1506 and by the European
Commission, Contract No. 1ST-1999-10033 ``Long Distance Photonic Quantum
Communication''.

\end{document}